\documentclass[aps,prl,reprint,superscriptaddress,twocolumn,showkeys,amsmath,amssymb]{revtex4-1}
\usepackage[english]{babel}
\usepackage{amsmath,amssymb,bbm,mathrsfs,bm,braket,color,graphicx,comment,amsfonts,dsfont}
\usepackage[colorlinks,citecolor=blue,urlcolor=blue]{hyperref}
\usepackage[mathscr]{euscript}

\newcommand{\Hc}{\mathrm{H.c.}}
\newcommand{\id}{\mathds{1}}

\newcommand{\pd}{{\phantom\dag}}

\newcommand{\tr}{\mathop{\mathrm{tr}}}

\begin{document}

\title{Localization counteracts decoherence in noisy Floquet topological chains}

\author{M.-T. Rieder}

\affiliation{Department of Condensed Matter Physics, Weizmann Institute of Science, Rehovot 7610001, Israel}

\author{L. M. Sieberer}

\affiliation{Department of Physics, University of California, Berkeley,
  California 94720, USA}

\affiliation{Institute for Theoretical Physics, University of Innsbruck, A-6020
  Innsbruck, Austria}

\affiliation{Institute for Quantum Optics and Quantum Information of the
  Austrian Academy of Sciences, A-6020 Innsbruck, Austria}

\author{M. H. Fischer}

\affiliation{Institute for Theoretical Physics, ETH Zurich, 8093 Zurich, Switzerland}

\author{I. C. Fulga}

\affiliation{IFW Dresden, Helmholtzstr.\ 20, 01069 Dresden, Germany}

\begin{abstract}
  The topological phases of periodically-driven, or Floquet systems, rely on a
  perfectly periodic modulation of system parameters in time. Even the smallest
  deviation from periodicity leads to decoherence, causing the boundary (end) states
  to leak into the system's bulk. Here, we show that in one dimension this decay
  of topologically protected end states depends fundamentally on the nature of
  the bulk states: a dispersive bulk results in an exponential decay, while a
  localized bulk slows the decay down to a diffusive process. The localization
  can be due to disorder, which remarkably counteracts decoherence even when it
  breaks the symmetry responsible for the topological protection. We derive this
  result analytically, using a novel, discrete-time Floquet-Lindblad formalism
  and confirm our findings with the help of numerical simulations. Our results
  are particularly relevant for experiments, where disorder can be tailored to protect
  Floquet topological phases from decoherence.
\end{abstract}

\maketitle

\textit{Introduction.}---Recently, Floquet systems have been established
as a novel paradigm of physics hosting a plethora of unique and fascinating
phenomena. Periodic modulation of a quantum system's Hamiltonian is not only a
powerful tool to engineer exotic, effectively static models, but even more
intriguingly it allows to realize novel phases of matter that do not have
time-independent counterparts. Experimental studies of such phases raise a
fundamental question: How robust are the new phenomena against unavoidable
imperfections in the implementation of the periodic driving? Here, we address
this question for the specific case of Floquet topological phases
\cite{Kitagawa2010, Jiang2011, Kitagawa2012, Rudner2013, Reichl2014,
  Carpentier2015, Nathan2015, Fulga2016, Leykam2016, Po2016,
  Morimoto2017, Quelle2017, Roy2017, Harper2017, Rudner2013, Titum2016, Liu2013,
  Maksymenko2017, Hu2015, Gao2016, Mukherjee2017, Maczewsky2017,Yao2017}. The unique
feature of these phases is that they have topologically protected boundary states
even if the bulk bands have vanishing topological invariants \cite{Rudner2013,Carpentier2015,Nathan2015,Yao2017} --- which is impossible in static systems. We ask, then, what the
fate of these boundary states is in the presence of imperfect or noisy driving.

Indeed, in any realistic experimental scenario, it is impossible to achieve
exact invariance under discrete time translations and thus, to conserve
quasienergy. For example, in realizations of Floquet topological
insulators with photonic waveguides \cite{Mukherjee2017, Maczewsky2017}, 
the propagation of light along the waveguide direction emulates time evolution
\cite{Longhi2009, Szameit2010}, and deviations from periodicity appear due to
fabrication defects. The latter cause the simultaneous presence of both static
(or quenched) disorder as well as temporal randomness, or noise. However, their
combined effect on the topological boundary states of Floquet systems has to our
knowledge never been addressed.

\begin{figure}[t!]
  \centering
    \includegraphics[width=.45\textwidth]{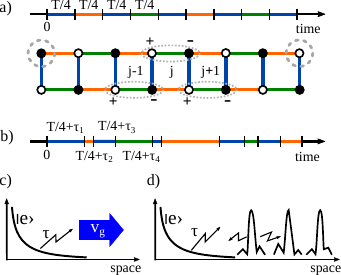}
    \caption{(Color online) Floquet chain with timing noise. (a) Driving protocol: $H_i$, $i = 1,...4$, is active for time interval $T_i = T/4$, enabling hopping along bonds as indicated by the colors. Sites are labeled according to their doublet $j$ (gray dashed ellipses) and their sublattice $s$ (empty/full circles). In a finite ladder of length $L$, two end sites marked by dashed circles have no doublet partner, $(0,-)$ and $(L,+)$, and host topologically protected end states. (b) Timing noise corresponds to fluctuations of the durations $T_i$ by random uncorrelated amounts $\tau_{i}$. (c) For a dispersive bulk, excitations out of the end state $\ket{e}$ due to the timing noise are carried away with a finite group velocity $v_g$, leading to an exponential decay. (d) For a localized bulk, excitations perform a random walk between localized states, resulting in a diffusive decay.}
  \label{fig:model}
\end{figure}

We study this problem for a Floquet topological system on a 1D ladder. It is subject to a piecewise-constant drive (see Fig.~\ref{fig:model}), consisting of steps during which the Hamiltonian is constant. We
use a simple but generic model for deviations from periodicity: a random
and uncorrelated-in-time duration of each driving step. Formally, this amounts
to timing noise in the piecewise-constant Hamiltonian of the system. We focus on
how imperfections in the periodic driving lead to the decay of the topologically
protected end state. While a system initialized in a Floquet eigenstate and
evolved with a noisy Hamiltonian will always lose memory of the
initial state, we show that the \emph{rate} of decay crucially depends on the
bulk's nature. If the bulk is dispersive the end state decays exponentially
in time. For localized bulk states it fades out diffusively.  We prove this by
deriving a discrete-time Floquet-Lindblad equation describing the noisy time
evolution of the system density matrix. We first apply it to a clean system with either a dispersive or a localized bulk, and then compare our results
to numerical simulations of a system with a disorder-localized bulk.

\textit{Model.}---Consider a model of spinless fermions on a 1D
ladder with sites labeled by $(j, s)$, where $j$ denotes a doublet of
sites and $s$ the sublattice index (see Fig.~\ref{fig:model}). The system is
subject to a periodic drive with frequency $\omega$ and period
$T = 2\pi/\omega$, consisting of four steps of equal duration $T/4$. In each
step, only hopping across certain bonds is allowed, as shown in
Fig.~\ref{fig:model}. The piecewise-constant Hamiltonian reads
\begin{equation}
  H(t) = - \sum_{\mu, \nu} J^\pd_{\mu\nu} (t) \left(c^{\dagger}_{\mu} c^\pd_{\nu} + \Hc \right),
  \label{eq:H0}
\end{equation}
where $c^{(\dag)}$ are fermionic annihilation (creation) operators and $\mu = \left( j, s \right)$. The hopping amplitudes are constant during each step $i=1,...4$
\begin{equation}
  J^\pd_{\mu \nu} (t) = J_{\mu \nu}^{i} \;\text{ for }\; (i-1)T/4 \leq t < iT/4 ,
  \label{eq:Js}
\end{equation}
where $J_{\mu \nu}^i = J$ for active bonds, and zero otherwise. The active bonds in steps $1$ and $3$ are the rungs of the ladder, connecting sites $(j,s)$ to $(j-s,-s)$. In step $2$, sites $(j,s)$ link to $(j-2s,-s)$ (across doublets) and in step $4$ to $(j,-s)$ (within doublets). This driving protocol is illustrated in Fig.~\ref{fig:model}.

The time evolution over one driving cycle is described by the Floquet operator,
$U_F = \mathsf{T} \exp\!\left(-i\int_0^T H(t)dt\right)$, where $\mathsf{T}$
denotes time-ordering and $\hbar=1$ throughout. Repeated
application of $U_F$ yields the stroboscopic evolution at multiples of the
period $T$.  By diagonalizing 
$U_F \ket{\alpha} = e^{-i T \varepsilon_{\alpha}} \ket{\alpha}$ we can describe
the system in terms of its Floquet eigenstates $\ket{\alpha}$ and their
associated quasienergies $\varepsilon_{\alpha}$. For the
Hamiltonian in Eq.~\eqref{eq:H0}, the Floquet operator factorizes as
$U_F = U_4 U_3 U_2 U_1$, where \cite{suppmat}
\begin{equation}
  \label{eq:Ui}
  U_i = P_i \left(\cos(\phi) - i \sin(\phi) \, H_i /J\right) P_i + Q_i.
\end{equation}
Here, $\phi = JT/4$ and $H_i$ denotes the Hamiltonian in
step $i$, i.e.,
$H_i = -\sum_{\mu \nu}J_{\nu\mu}^i \left( c^\dagger_\mu c^\pd_\nu + \Hc \right)$. The
operators $P_i$ are projectors onto all lattice sites appearing in $H_i$ and $Q_i = \id - P_i$. Note that for $i=1,3$ all lattice
sites are involved and thus $P_i = \id$, whereas in step $2$ the lattice sites
$\left( 1, + \right)$ and $\left( L-1,- \right)$ are excluded and, similarly, in step $4$ the sites
$\left( 0, - \right), \left( L, + \right)$.

The properties of Floquet eigenstates are governed by the phase
$\phi$. A special case occurs at the \emph{resonant driving point}
$\phi = \pi/2$, where the otherwise dispersive bulk is fine-tuned to form a
localized flat band: in every step of the driving protocol, a particle is fully
transferred between two lattice sites, acquiring a phase of $\pi/2$. As
such, particles initialized in the Floquet eigenstates
$\ket{j,s} = c^{\dagger}_{j,s}\ket{0}$ encircle the $j$-th plaquette in
opposite directions for $s=\pm$ during each period. In one period, each bulk state accumulates a phase of $2 \pi$ resulting in a flat band at
quasienergy $\varepsilon = 0$. At the chain's ends, however, there are two
states, $\ket{0,-}$ and $\ket{L,+}$, without doublet partners and
therefore skip two steps. This results in an accumulated phase of $\pi$ and a quasienergy $\varepsilon = \omega/2$.

Away from resonant driving, i.e., $\phi = \pi/2 + \delta \phi$, the bulk band becomes dispersive with a bandwidth $\propto \delta\phi$ and the bulk states $\ket{b}$ are delocalized over the entire chain. 
The left and right end states $\ket{e_{l/r}}$ at $\varepsilon=\omega/2$ are 
topologically protected and remain exponentially localized on the boundaries, due to the ladder's sublattice symmetry. If it is broken, e.g. by an onsite potential, the end states can be pushed away from $\omega/2$ into the system's bulk~\cite{suppmat}. Experimentally, fine-tuning the system to resonant driving is unrealistic \cite{Mukherjee2017, Maczewsky2017}, suggesting the dispersive bulk as the generic case. The resonant point still provides a natural example of a localized bulk in our model and is a good starting point to study the effect of noise. 

\textit{Timing noise and Floquet-Lindblad equation.}--- We introduce deviations
from the perfectly-periodic time dependence of $H(t)$, Eq.~\eqref{eq:H0}, through timing noise in the driving protocol. The duration of step $i$ in
the $n$-th cycle is no longer $T/4$, but $T/4 + \tau_{ni}$. Here, $\tau_{ni}$ are random numbers from a Gaussian distribution with zero
mean, $\overline{\tau_{ni}} = 0$, and fluctuations
$\overline{\tau_{ni} \tau_{n'i'}} = \tau^2 \delta_{nn'} \delta_{ii'}$, i.e., the
deviations are uncorrelated between steps and cycles. We consider a weak noise strength $\tau$ ($\tau J \ll 1$). The dynamics is
then described by a ``noisy Floquet operator''
$U_{F, n} = U_{4, n} U_{3, n} U_{2 ,n} U_{1, n}$, with $U_{i, n}$ as in Eq.~\eqref{eq:Ui} with $\phi \to \phi + J \tau_{ni}$, such that the state of the system after $n$ noisy driving cycles is $\ket{\psi_n} = U_{F, n} \dotsb U_{F, 1} \ket{\psi_0}$. Since only the product
$J \tau_{ni}$ enters the time evolution, timing noise is equivalent to
fluctuations of the hopping amplitude $J$. 

The noise-averaged effective description of the system's dynamics is captured in
a Lindblad-like equation. To derive it, we first note that the expectation value
of any observable $O$ at time $nT$ can be written as
$\overline{\langle O_n \rangle} = \overline{\braket{\psi_n | O | \psi_n}} =
\tr(O \rho_n)$,
with the density matrix $\rho_n = \overline{\ket{\psi_n} \bra{\psi_n}}$ and a
noise average $\overline{\cdots}$. As the timing noise is uncorrelated for
different cycles, we can derive a stroboscopic evolution equation for the
density matrix over one cycle
$\rho_{n+1} = \overline{U_{F, n + 1} \rho_n U_{F, n + 1}^\dagger}$. Expanding
$U_{F, n}$ to second order in the noise strength $\tau$ and performing the noise
average we find that the evolution equation takes the form of a discrete-time
Floquet-Lindblad equation~\cite{suppmat}:
\begin{align}
  \label{eq:lindblad}  
  \rho_{n+1} =  U_F \biggl( \rho_{n}  
  + \tau^2 \sum_i \mathcal{D}[L_i] \rho_n \biggr)
  U_F^\dagger .
\end{align}
$\mathcal{D}[L] \rho = L \rho L - \frac{1}{2} \left\{ L^2, \rho \right\}$ is the
so-called dissipator, and the self-adjoint quantum jump operators are 
$L_1 = H_1 $, $L_2 = U_1^\dagger H_2 U_1$,
$L_3 = U_1^\dagger U_2^\dagger H_3 U_2 U_1$, and
$L_4 = U_1^\dagger U_2^\dagger U_3^\dagger H_4 U_3 U_2 U_1$. As in the
continuous-time Lindblad equation (see, e.g.,~\cite{Gardiner2000}), dissipation
originates from perturbatively eliminating fluctuations that are uncorrelated on
the characteristic time scale of the system dynamics --- the main difference
being that usually these fluctuations correspond to additional quantum degrees
of freedom, while here they are induced by noise.

The influence of noise on the stroboscopic dynamics can be seen best by
expanding the density matrix in a basis of Floquet eigenstates. For simplicity,
we assume a semi-infinite chain with left end state $\ket{e} \equiv \ket{e_l}$
and bulk basis states $\ket{b}$. We study the noise-induced decay of the left end state occupation, $\rho^e_n = \braket{e | \rho_n | e}$, starting
from the pure state $\rho_0 = \ket{e} \bra{e}$. The coherent evolution described
by $U_F$ conserves the occupations of Floquet eigenstates, while dissipation
leads to transitions due to finite matrix elements $\braket{b | L_i | e}$.

We can quantify this decay close to resonant driving, where $\phi = \pi/2 + \delta \phi$ with $\delta \phi \ll 1$. There, up to corrections $O(\delta \phi)$, the matrix elements of the operators $L_i$ on the end
states take the simple forms $\braket{e|L_i|e} = 0$,
$\braket{e|L_i^2|e} = J^2$ for $i=1,3$ and $0$ for $i=2,4$, and
$\braket{b | L_i^2 | e} = 0$~\cite{suppmat}. We thus find
\begin{align}
\label{eq:rho_edge}
  \rho^e_{n+1} = \left( 1- 2 J^2 \tau^2 \right) \rho^e_{n} 
  + \tau^2 \sum_i \sum_{bb'} \braket{e|L_i|b} \braket{b'|L_i|e} \rho^{bb'}_n ,
\end{align}
where $\rho^{bb'}_n =\braket{b|\rho_n|b'}$. While the above-mentioned matrix
elements are well-behaved for $\delta \phi \to 0$ and corrections due to finite
$\delta \phi$ can safely be disregarded, the sum on the right-hand side of
Eq.~\eqref{eq:rho_edge} is non-analytic. This results in fundamentally different
behavior for the limit $\delta \phi \to 0$ and at
resonant driving $\delta \phi =0$ itself, as we discuss in the following.

\textit{Exponential Decay.}---For $\delta \phi \to 0$
the off-diagonal terms in the density matrix oscillate strongly, reducing
the dynamics to an exponential decay. To see this, note that the bulk states
are dispersive with quasienergies much smaller than the driving
frequency $\epsilon_b \ll \omega$, but---crucially---nonzero. The end state is
exponentially localized, while the bulk states are spread out over the chain. Thus, the matrix elements of the quasilocal operators $L_i$ connecting
end state and bulk decay with the system size:
$\braket{e|L_i|b} \sim J/\sqrt{L}$. Further, due to noise the off-diagonal
terms of the bulk's density matrix $\rho_n^{bb'}$ are populated with a rate
$\sim J^2\tau^2/L$ per period, but acquire phases from the coherent time
evolution at a rate proportional to 
$(\epsilon_b-\epsilon_{b'}) \sim \delta \phi$. 
For system sizes $L \gg J^2\tau^2/\delta \phi$ these off-diagonal terms dephase
faster than they are populated, and render the density matrix approximately
diagonal at long times. Assuming the bulk states are populated evenly from
excitations out of the end state, the diagonals are $\rho_n^{bb} = O(1/L)$. The
sum in Eq.~\eqref{eq:rho_edge} thus scales with $1/L$ and can be neglected for large enough systems, resulting in an
exponential decay with a rate $2 J^2 \tau^2$:
\begin{equation}
  \rho^e_{n+1} = \left( 1- 2 J^2\tau^2\right) \rho^e_n + O(1/L) \,. \label{eq:exp_decay}
\end{equation}
A qualitatively similar result was found for decoherence
in quantum walks \cite{Groh2016}.

Heuristically, the finite bandwidth of the bulk band allows excitations from the
end state into the bulk to move away from the edge fast enough to not lead to
blocking or backflow. This is the origin of the fast (exponential)
decay. Conversely, for a localized bulk, we expect a qualitatively different
behavior. This can be seen close to resonant driving, i.e.,
$ J^2\tau^2 / \delta\phi \gg L$, which we give as a bound on tolerable
deviations from resonant driving in experiments. Alternatively and more
realistically, the bulk can be localized by disorder. We show in the following
how both scenarios lead to diffusive decay of the end-state occupation.

\textit{Diffusive Decay.}---At resonant driving, the Floquet-Lindblad equation can be analyzed exactly and simplifies into a diffusion equation at long times. We start noting that due to the full degeneracy of the bulk band, the bulk basis is not fixed. 
In the real-space basis $\ket{b}=\ket{j,s}$, we can treat the evolution equation exactly for any state in the ladder. Now, the off-diagonal elements of $\rho_n$ are not populated from the diagonal
\cite{suppmat}
and we disregard them throughout the calculation as we consider in a diagonal initial state $\rho_0 = \ket{e} \bra{e}$. The diagonal elements, $\rho_n^{j,s} = \braket{j,s|\rho_n|j,s}$, evolve under
the noisy periodic driving as
\begin{equation}
\begin{split}
  \rho_{n+1}^{j,s} = & \left( 1 - 4 J^2\tau^2 \right) \rho_n^{j,s} 
  + \\ & J^2\tau^2 \left( \rho_n^{j-s,-s} + 2 \rho_n^{j, -s} + \rho_n^{j+s,-s} \right).
\end{split}
\end{equation}
The total occupation of the doublet $j$ is given by the sum
$\rho^j_n = \rho^{j,+}_n + \rho^{j,-}_n$. It obeys a discrete diffusion
equation, while the relative occupation of the two sublattice sites,
$\sigma^j_n = \rho^{j,+}_n - \rho^{j,-}_n$, is exponentially suppressed:
\begin{align}
  \label{eq:diff_eq}
  D_t \rho^j_n & = \Delta  D_x^2 \rho^j_n, \\
  D_t \sigma^j_n & = \left(-8 J^2\tau^2 -\frac{J^2\tau^2}{T} D_x^2 \right)
                   \sigma^j_n.
\end{align}
The diffusion coefficient is $\Delta ={J^2\tau^2 a^2/ T}$ and we introduced
the discrete time derivative $D_t f_n = (f_{n+1}-f_n)/T$ and second order
spatial derivative $D_x^2 f^j = (f^{j+1} - 2 f^j + f^{j-1})/a^2$ with lattice
constant $a$. While this diffusion equation describes the dynamics of an arbitrary bulk state $\ket{j, s}$, we emphasize that in the lattice-site basis bulk and end states can be treated on an equal footing. Thus, the wave function of a particle initialized in an end state $\ket{e} = \ket{0, -}$ spreads diffusively into the bulk, assuming a Gaussian profile of width $\sim \sqrt{n T}$.

\textit{Disorder.}---While the drastic slow down of the end state decay seems
out of reach at a fine-tuned point in parameter space, its cause is ultimately a
localized bulk band. This suggests that adding disorder should have a similar
effect and in the following we argue that it indeed results in a diffusive
process, effectively protecting the end state against decay. 
We consider
disorder in either the hopping matrix elements, allowing for site-dependent fluctuations $ J_{\nu\mu}^i + \delta j_{\mu \nu}$, or in the on-site potential, adding a random
chemical potential with strength $v_{\mu}$ to the Hamiltonian. 
We take
$\delta j_{\mu \nu}$ and $v_{\mu}$ to be uncorrelated and uniformly distributed
in the interval $[-V, V]$, with $V$ the disorder strength. In the presence of
disorder, Floquet eigenstates $\ket{l}$ are localized on a length scale
$\xi$. They are labeled according to their center of mass along the ladder,
such that $\ket{l}$ can be thought of as being the left nearest neighbor of
$\ket{l+1}$. As observed above, the off-diagonal density-matrix elements typically play
no role in the long-time behavior of the system. Therefore,
Eq.~\eqref{eq:lindblad} can again be reduced to a classical master equation
\cite{suppmat},
\begin{equation}
  \rho_{n+1}^l = \rho_n^l + \sum_{l' \neq l} W_{l' \to l} \left(\rho_n^{l'} - \rho_n^l \right),
\end{equation}
where $\rho^l_n = \braket{l|\rho_n|l}$ and
$W_{l' \to l} = \tau^2 \sum_i \left|\braket{l'|L_i|l}\right|^2$. The operators
$L_i$ are defined as in Eq.~\eqref{eq:lindblad}. Since these operators are
quasilocal, the transition probabilities $W_{l' \to l}$ only connect nearby
states. Thus, the time evolution of the density matrix is reduced to a random
walk with disordered transition probabilities, and as such must be
diffusive \cite{Hughes1995}. Note that this behavior only requires a localized bulk, independently of the type of disorder. As such, on-site disorder will also lead to an increased robustness of the end state, even though it breaks the sublattice symmetry required for its protection. For small strengths of the chemical potential disorder $V$, the end modes will shift from the quasienergy $\omega/2$ by an amount $\sim V$, but will still decay diffusively. \footnote{Note that for sufficiently small disorder strength the end state are still well separated in quasienergy from the bulk states. As such, they are still well defined, even if they are not topologically protected. Such an unexpected stability of Floquet-topological phases against symmetry-breaking perturbations has been reported also in previous work \cite{Balabanov2017}.}

\begin{figure}
  \centering
    \includegraphics[width=.45\textwidth]{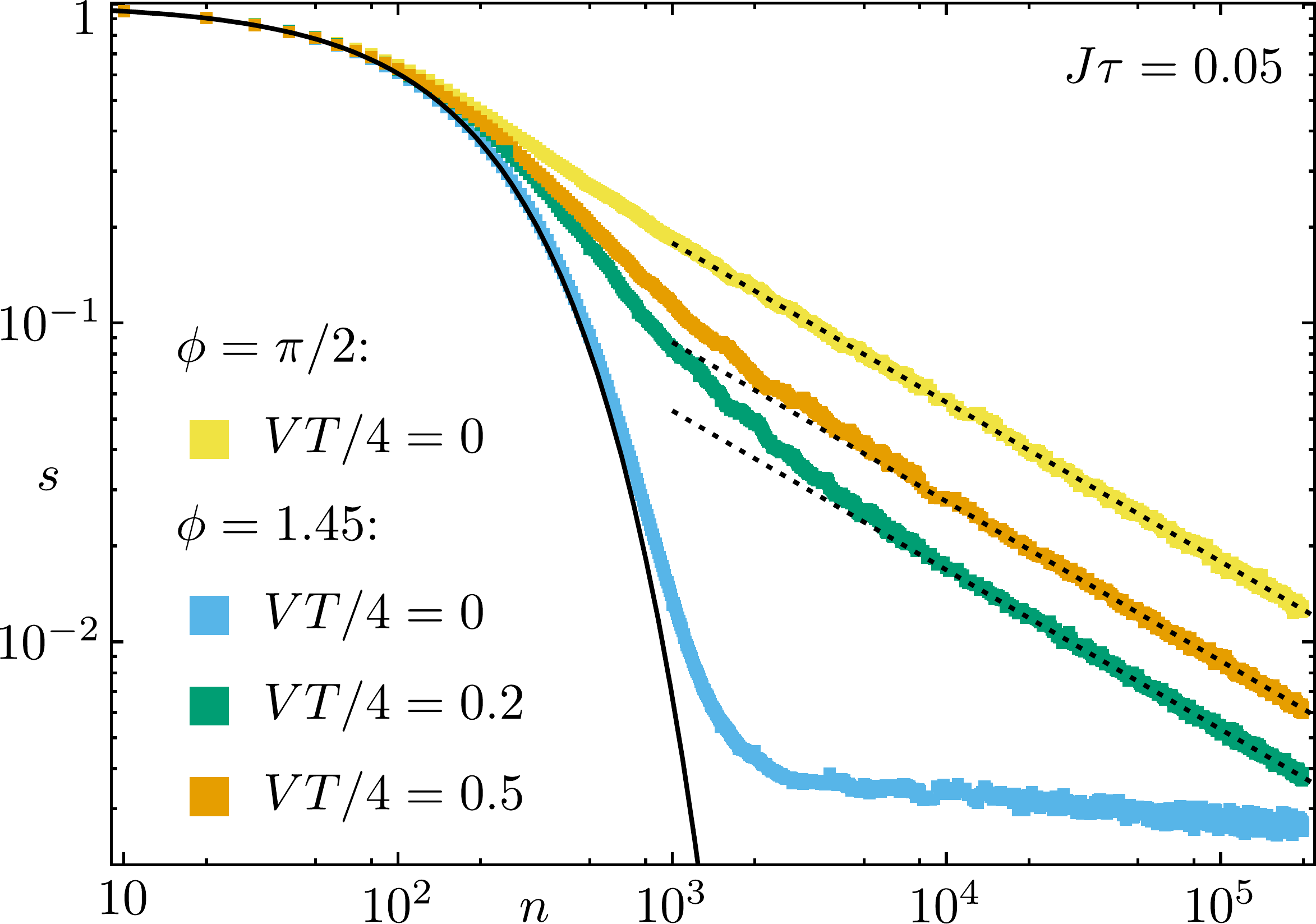}
    \caption{(Color online) End state survival probability
      $s = \overline{|\braket{e | \psi_n}|^2}$ with driving cycles $n$. A clean,
      dispersive system (blue) leads to exponential decay with finite-size
      saturation at long times. A localized bulk shows diffusive decay at long
      times, both for a clean system at resonant driving (yellow) and a
      disordered system with onsite disorder strength $V$ (orange and
      green). For comparison, the black solid (dashed) line indicates purely
      exponential decay with rate $2J^2\tau^2$ (diffusive decay $\sim t^{-1/2}$
      where $t = n T$). The blue (yellow) curve is averaged over $2000$ (5000)
      noise realizations, while the green and orange curves are averaged over
      $4000$ simultaneous disorder-noise configurations.}
  \label{fig:1d_disorder}
\end{figure}

\textit{Numerics.}---To corroborate our analytical results, we numerically
integrated the Schr\"odinger equation with the driven and noisy
Hamiltonian~\eqref{eq:H0} for a ladder with $200$ rungs.  Figure
\ref{fig:1d_disorder} shows the survival probability of the end state
$s = \overline{|\braket{e | \psi_n}|^2}$ (averaged over noise realizations) for
a single particle initialized at the edge, $\ket{\psi_0} = \ket{e}$. As
expected, we find an exponential decay of the end state at a rate
$2 J^2\tau^2$ [cf.~Eq.~\eqref{eq:exp_decay}] for a dispersive bulk, while
localized bulk states lead to a much slower, diffusive decay --- irrespective of
whether localization is due to fine-tuning to resonant driving in a clean system
or disorder. We have numerically studied both hopping and on-site
disorder \cite{suppmat}.

\textit{Conclusions.}---We studied the combined effect of quenched disorder and
noise in a 1D Floquet topological phase. We showed analytically and numerically
that a dispersive bulk causes an exponential decay of the end mode due to noise,
while a localized bulk is associated with a qualitatively slower, diffusive
decay. While we focused on a particular model of timing noise, we expect our
results to hold also for other forms of temporal disorder which is
uncorrelated on a time scale set by the driving period. The formalism we
developed provides a comprehensive framework to address timing noise in Floquet
systems with piecewise constant driving, including periodically ``kicked''
(non-topological) systems which have been the focus of recent
experimental~\cite{Oskay2003,Bitter2016,Bitter2017} and
numerical~\cite{Cadez2017} studies. For future work, it will serve as a ground
to extend the Floquet-Lindblad formalism to smooth drivings with correlated
noise, or study the interplay of quenched disorder and noise in higher dimensions.
Moreover, our approach of studying the
noise-averaged stroboscopic evolution of the density matrix can be also be
applied to models with strong noise, for which the stroboscopic evolution does
not take the Floquet-Lindblad form~\eqref{eq:lindblad}. It is an intriguing
question whether the steady states of such models could have interesting
topological properties.

Relying only on localization and not on system details, our results hold for
many 1D topological phases, such as the SSH \cite{Su1979} or the Kitaev
\cite{Kitaev2001} chain, and are highly relevant to realizations of Floquet
topological phases \cite{Kitagawa2010, Kitagawa2012, Hu2015, Gao2016,
  Mukherjee2017, Maczewsky2017}. In fact, we showed that purposely including
quenched disorder can increase the robustness of a topological boundary against
unavoidable losses in an experimental setup. This increase can be by orders of
magnitude at long times, and occurs even when the disorder breaks the symmetries
required for topological protection.  Our results can be readily tested in
experiments, since the 1D ladder we consider is a simplified version of the
already experimentally realized 2D anomalous Floquet topological phases
\cite{Mukherjee2017, Maczewsky2017}.

\begin{acknowledgments}
  We thank E.~Altman and C.~Roberto for insightful discussions, as well as Ulrike Nitzsche for technical assistance. LMS and MTR
  thank KITP for hospitality. Further, MTR, LMS, and MHF owe thanks to the WIS
  trainee programm for its support during the early stages of this work. MTR
  ackowledges support by the Alexander von Humboldt Foundation and LMS
  acknowledges support by the ERC through the synergy grant UQUAM.
\end{acknowledgments}

\bibliography{bibliography}

\end{document}


\title{Supplemental Material: Localization counteracts decoherence in noisy Floquet topological chains}

\author{M.-T. Rieder}

\affiliation{Department of Condensed Matter Physics, Weizmann Institute of Science, Rehovot 7610001, Israel}

\author{L. M. Sieberer}

\affiliation{Department of Physics, University of California, Berkeley,
  California 94720, USA}

\affiliation{Institute for Theoretical Physics, University of Innsbruck, A-6020
  Innsbruck, Austria}

\affiliation{Institute for Quantum Optics and Quantum Information of the
  Austrian Academy of Sciences, A-6020 Innsbruck, Austria}

\author{M. H. Fischer}

\affiliation{Institute for Theoretical Physics, ETH Zurich, 8093 Zurich, Switzerland}

\author{I. C. Fulga}

\affiliation{IFW Dresden, Helmholtzstr. 20, 01069 Dresden, Germany}

\begin{abstract}
In this Supplemental Material, we provide a detailed discussion of
the topological properties of the model considered in the
main text, the effect of adding disorder to this model, and
details of the derivation of the Floquet-Lindblad equation, both for stroboscopically driven systems with timing noise in general and the specific model we consider.
\end{abstract}

\maketitle

\section{Floquet topological phase of the 1D ladder}

\begin{figure}[h]
 \includegraphics[width=0.4\textwidth]{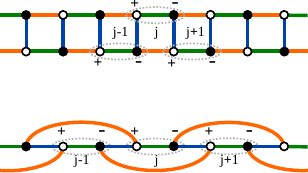}
 \caption{For an infinite 1D ladder (top), we can determine the momentum-space Hamiltonian by rearranging the site doublets denoted by $j$ such that they become the unit cells of an infinite chain (bottom). The hoppings on the rungs of the ladder (blue) then connect neighboring unit cells. Half of the hoppings on the legs of the ladder (green) connect sites within a unit cell, while the other half (orange) become next-nearest-neighbor hoppings. \label{fig:straighten}}
\end{figure}

The 1D ladder model introduced in the main text is an example of an anomalous Floquet topological phase, one in which topologically protected end modes appear even though the bulk bands have vanishing topological invariants. To see this, we write the Hamiltonian and the Floquet operator in momentum space by ``straightening out'' the ladder, as shown in Fig.~\ref{fig:straighten}.

The resulting $2\times2$ momentum-space Hamitonian reads
\begin{equation}\label{eq:bulkH}
 H(k, t)=\begin{pmatrix}
    0 & q(k, t) \\
    q^*(k, t) & 0 \\
   \end{pmatrix},
\end{equation}
where $q(k, t)=J^4(t)+J^{1,3}(t) e^{i k}+J^2(t) e^{2ik}$. As in the main text, we denote by $J^{1,3}$ the hoppings on the rungs of the ladder, by $J^4$ the hopping within doublets, and by $J^2$ the one across doublets. 

Before imposing a periodic modulation of the $J^i$, we examine the symmetries of the Hamiltonian Eq.~\eqref{eq:bulkH} in the time-independent case. Since we consider real hopping amplitudes, the Hamiltonian Eq.~\eqref{eq:bulkH} exhibits time-reversal ${\cal T}$, particle-hole ${\cal P}$, as well as sublattice $\Gamma$ symmetries. Introducing Pauli matrices $\sigma_{x, y, z}$ acting on the sublattice degree of
freedom, the symmetry operators are $\Gamma = \sigma_z$, ${\cal T} = {\cal K}$, and ${\cal P} = \sigma_z {\cal K}$, with ${\cal K}$ complex conjugation, such that
\begin{equation}\label{eq:bulkHsym}
\begin{split}
 H(k) & = H^*(-k) \\
 H(k) & = - \sigma_z H^*(-k) \sigma_z \\
 H(k) & =  -\sigma_z H(k) \sigma_z \\
\end{split}
\end{equation}
 Notice
that ${\cal T}^2={\cal P}^2=+1$, such that $H(k)$ belongs to class BDI in the Altland-Zirnbauer classification \cite{Altland1997}.

\begin{figure}[h]
 \includegraphics[width=0.3\textwidth]{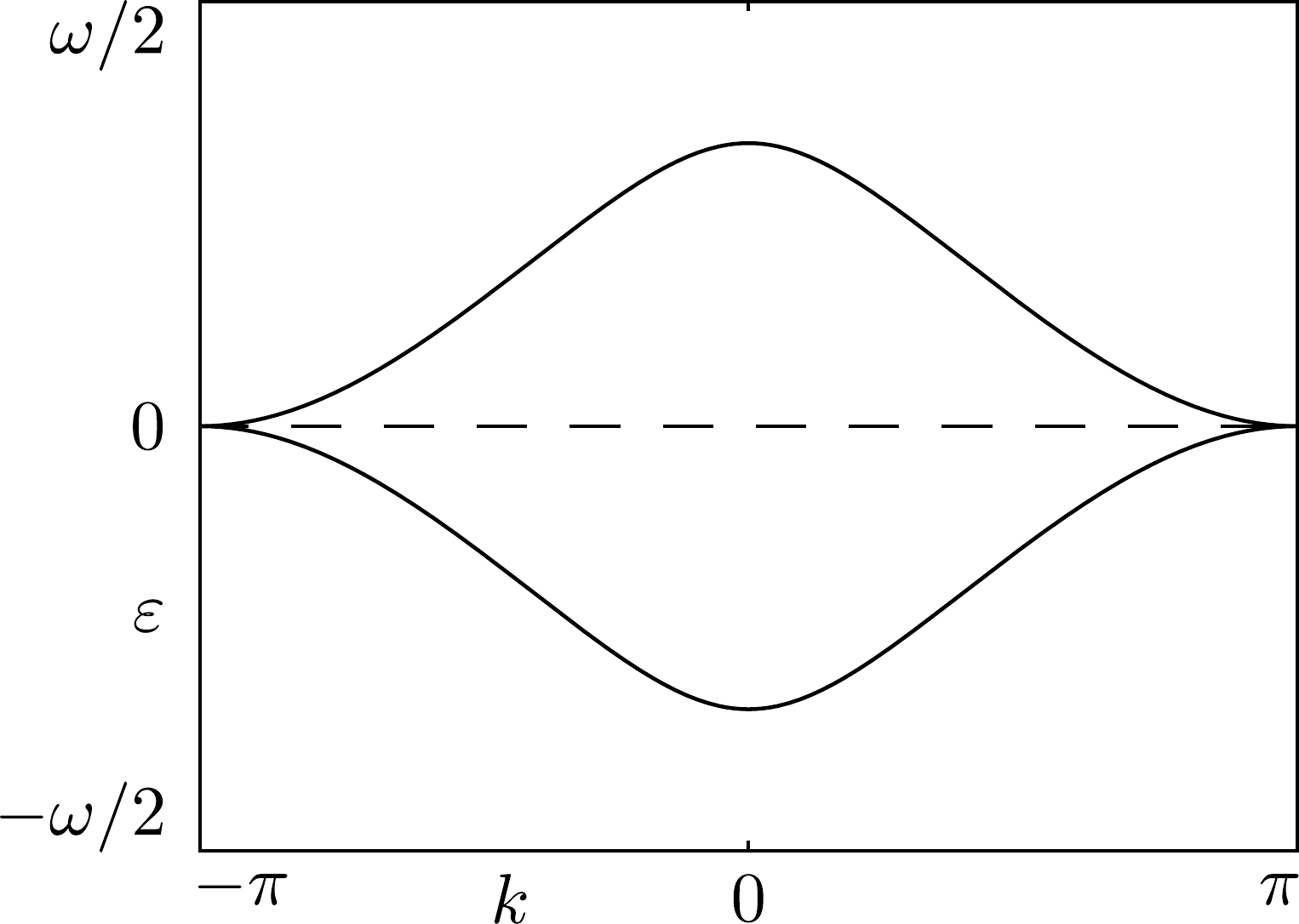}
 \caption{Bandstructure of the Floquet operator Eq.~\eqref{eq:Floquetbulk} for $JT/4=\pi/2$ (dashed) and $JT/4=\pi/3$ (solid). At the resonant driving point the bulk bands are flat, whereas they disperse otherwise and form a parabolic touching point at $k=\pi$ and $\varepsilon=0$.\label{fig:bulkbands}}
\end{figure}

Next, we consider the case of time-periodic hopping amplitudes. As discussed in the main text, during each of the four steps of the driving protocol only
one of the hoppings is active, such that the hopping $J^i=J$ for $(i-1)T/4 \leq t < iT/4$ and all other hoppings vanish.
Owing to the simple form of the Hamiltonian and of the driving protocol, the Floquet operator takes the form
\begin{equation}\label{eq:Floquetbulk}
U_{0\rightarrow T}(k)=U_F(k) = e^{-i \frac{T}{4} H_4(k)}e^{-i \frac{T}{4} H_3(k)}e^{-i \frac{T}{4} H_2(k)}e^{-i \frac{T}{4} H_1(k)},
\end{equation}
where $U_{0\rightarrow T}(k)$ denotes the time-evolution operator from time $t=0$ to $t=T$. The Floquet operator
can be computed exactly, although its form is involved for generic parameter values. At the resonant driving point however, when the strength of the active hopping $J$ is such that $JT/4=\pi/2$, the bulk Floquet operator becomes the identity operator $U_F(k)=\mathds{1}$. As such, the bulk Floquet bands are flat and positioned at zero quasienergy. In addition, this form of $U_F$ explains the anomalous nature of the topological phase: since Floquet eigenstates become momentum-independent at the resonant driving point, they must be topologically trivial. Away from the fine-tuned value $JT/4=\pi/2$, bulk bands are dispersive and show a parabolic band touching point at $k=\pi$, as shown in Fig.~\ref{fig:bulkbands}.

As discussed in the main text, the 1D ladder hosts a topological Floquet phase, such that end states appear in a finite system despite the topologically trivial bands. The presence of end states at quasienergies $\varepsilon=\pm\omega/2$ can be deduced directly from the real-space Floquet system at the resonant driving point, but remain pinned to these quasienergies also away from it. This behavior is linked to the symmetries of the Floquet operator. To determine them, it is convenient to choose a different origin of time when computing the Floquet operator and go to a so called \emph{symmetric time frame} by defining:
\begin{equation}\label{eq:Floquetsym}
U_{\frac{3T}{8}\rightarrow T+\frac{3T}{8}}(k)=\widetilde{U}_F(k) = e^{-i \frac{T}{8} H_2(k)}e^{-i \frac{T}{4} H_1(k)}e^{-i \frac{T}{4} H_4(k)}e^{-i \frac{T}{4} H_3(k)}e^{-i \frac{T}{8} H_2(k)}.
\end{equation}

For the time-shifted Floquet operator, time-reversal, particle-hole, and sublattice symmetry take the respective forms:
\begin{gather}\label{eq:FTRS}
 {\widetilde{U}}^\pd_F(k) = {\widetilde{U}}^T_F(-k), \\
\label{eq:FPHS}
 {\widetilde{U}}^\pd_F(k) = \sigma_z {\widetilde{U}}^*_F(-k) \sigma_z, \\
\label{eq:FSLS}
 {\widetilde{U}}^\pd_F(k) = \sigma_z {\widetilde{U}}^\dag_F(k) \sigma_z,
\end{gather}
where the superscript $T$ denotes transposition.
Note that out of these three constraints only particle-hole symmetry does not reverse the order of the product in Eq.~\eqref{eq:Floquetsym}, since it relates the Floquet operator to its complex conjugate. This means that the same constraint Eq.~\eqref{eq:FPHS} holds for all choices of initial time, and in particular also for the Floquet operator $U_F(k)$ of Eq.~\eqref{eq:Floquetbulk}, while time-reversal and sublattice symmetries acquire complicated, $k$-dependent forms.
As such, we will focus in the following on the un-shifted Floquet operator $U_F(k)$ and on the role played by particle-hole symmetry in protecting its end states.
Due to this symmetry, momentum eigenstates must come in pairs, related by $\varepsilon\to-\varepsilon$ and $k\to-k$, as seen in Fig.~\ref{fig:bulkbands}. In real space, particle-hole symmetry relates states at the same position but opposite values of quasienergy, explaining the robustness of the end modes. Even if the system is perturbed away from resonant driving, the end modes cannot couple due to the bulk gap so they cannot shift in quasienergy away from the particle-hole symmetric $\varepsilon=\omega/2=-\omega/2$.

\begin{figure}[h]
 \includegraphics[width=0.6\textwidth]{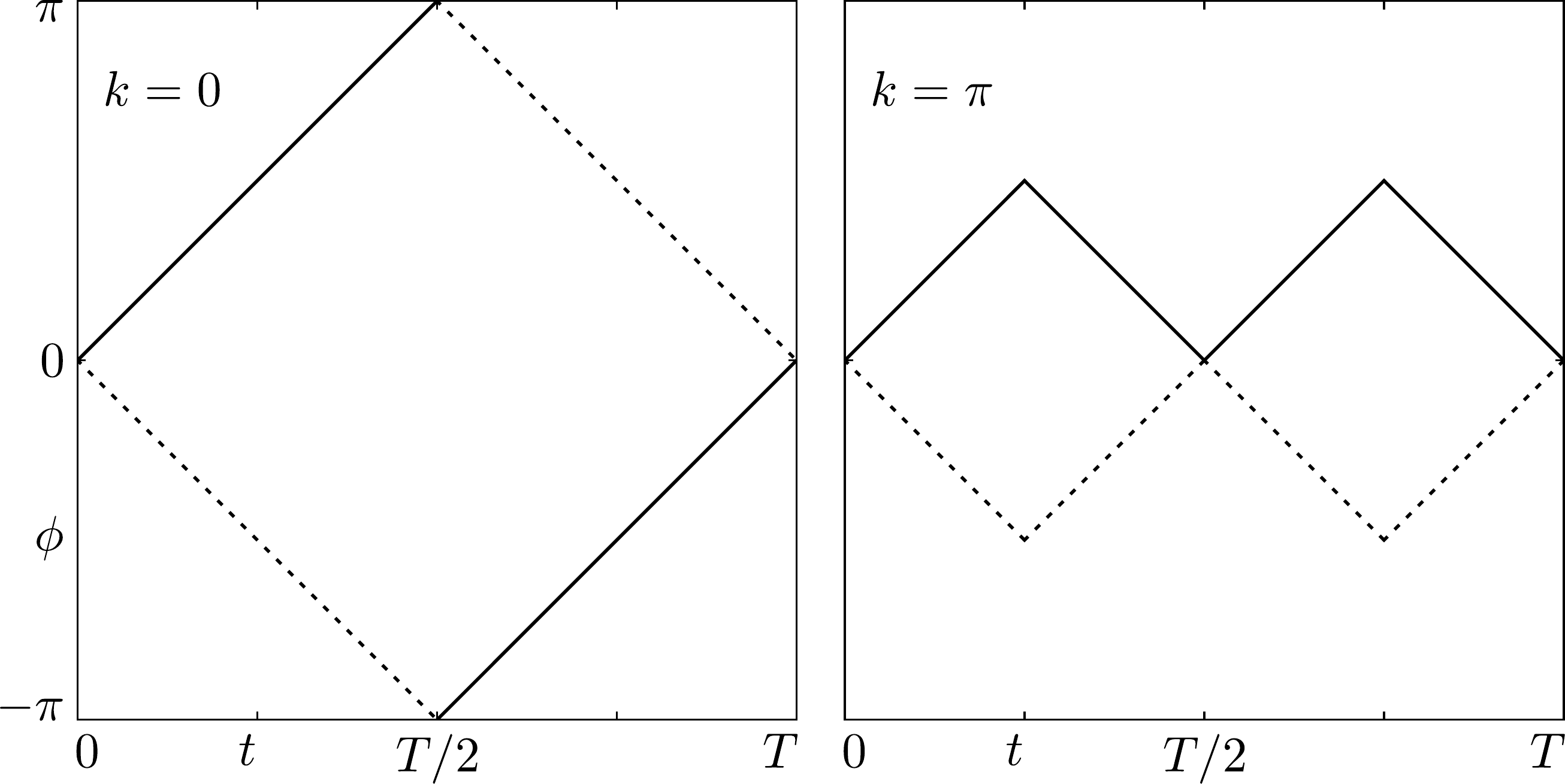}
 \caption{Eigenphases of $U_{0\rightarrow t}(k=0)$ (left) and $U_{0\rightarrow t}(k=\pi)$ (right) as $t$ is advanced from $0$ to $T$, using $JT/4=\pi/2$. Pairs of opposite eigenphases are denoted by solid and dashed lines. On the left, eigenphases are monotonic, since all $J^i$ enter the Hamiltonian with the same sign for $k=0$. On the right, the eigenphases reverse direction in each driving step, since $J^{1,3}$ and $J^{2,4}$ have opposite signs for $k=\pi$. The total number of $\pi$-crossings is odd, signaling a topologically non-trivial phase.\label{fig:topcrossing}}
\end{figure}

Since the driven system shows particle-hole, time-reversal, as well as sublattice symmetries, it belongs to symmetry class BDI \cite{Altland1997}, which was shown to have a $\mathbb{Z}$ topological classification in one dimension. To determine the topological invariant responsible for the presence of end states, we use the method of Ref.~\cite{Jiang2011}, which is based on evaluating the time-evolution operator at all times during the driving cycle. Writing
\begin{equation}
 U_{0\rightarrow t}(k) = \mathsf{T} \exp \left( {-i \int_0^t dt' \, H(k,t')} \right),
\end{equation}
the topological invariant can be determined as the total number of times the eigenphases of both $U_{0\rightarrow t}(k=0)$ and $U_{0\rightarrow t}(k=\pi)$ cross $\pi$ in the interval $t\in[0,T]$. Due to the simple form of the piecewise constant Hamiltonian Eq.~\eqref{eq:bulkH}, these eigenphases can be evaluated exactly, since $H(k=0)=(J^4+J^{1,3}+J^2)\sigma_x$ and $H(k=\pi)=(J^4-J^{1,3}+J^2)\sigma_x$. Therefore, the two eigenphases of $U_{0\rightarrow t}(k=0,\pi)$ are opposite and increase or decrease linearly in each driving step, depending on whether $J^i$ enters the Hamiltonian with a positive or negative sign. At resonant driving, for $k=0$ the eigenphases are monotonic as a function of $t$ and show a single crossing at $\pi$, when $t=T/2$. For $k=\pi$, the eigenphases reverse direction in steps 1 and 3 as compared to steps 2 and 4, such that they never cross $\pi$. As such, the total number of $\pi$-crossings is nonzero, as shown in Fig.~\ref{fig:topcrossing}, implying a topologically non-trivial phase.

\section{Timing noise in Floquet systems with piecewise constant driving}
 \label{sec:timing-noise-floquet-pc-drive}

Here we derive the discrete-time Floquet-Lindblad equation (FLE) that describes the
time evolution of periodically driven systems with timing noise. First we
consider the simplest case of binary driving (or, equivalently, a system
subject to periodic kicks), and then generalize our results to driving cycles
comprising an arbitrary number of steps. We discuss the conditions under
which the FLE can be reduced to a classical master equation. Then, we apply
the general formalism to the model system considered in the main text.
 
\subsection{Binary piecewise constant driving}
\label{sec:binary-pc-driv}

Let us begin with the simplest case, which is a binary and piecewise constant
driving protocol. We first discuss perfectly periodic driving, and then
modifications due to timing noise. Without noise, a driving cycle consists of
the following two steps: first, the Hamiltonian $H_1$ is applied for a time
$T_1$, and then another Hamiltonian $H_2$ is applied for a time $T_2$. The full
duration of the driving cycle is thus $T = T_1 + T_2$, and the Floquet operator
$U_F$, which describes the evolution of the system during one period, is given
by
\begin{equation}
  \label{eq:UF_binary}
  U_F = \mathsf{T} e^{-i \int_0^T dt \, H(t)} = U_2 U_1 = e^{-i T_2 H_2} e^{-i
    T_1 H_1}.
\end{equation}
Here, $\mathsf{T}$ denotes time ordering, and $H(t)$ is the time-dependent
Hamiltonian 
\begin{equation}
  H(t) =
  \begin{cases}
    H_1, & n T \leq t < n T + T_1, \\
    H_2, & n T + T_1 \leq t < \left( n + 1 \right) T.
  \end{cases}
\end{equation}
The integer $n$ counts the number of driving cycles. A Floquet operator with the
same form as in Eq.~\eqref{eq:UF_binary} arises in periodically ``kicked''
systems, in which $H_2$ is applied as an instantaneous pulse at multiples of the
period $T$. Such a scenario is described by the following time-dependent
Hamiltonian:
\begin{equation}
  \label{eq:H_kicked}
  H(t) = H_1 + \lambda \sum_{n \in \N} \delta(t - n T) H_2.
\end{equation}
For this type of driving, the noiseless evolution is given by the Floquet
operator $U_F = e^{-i \lambda H_2} e^{-i T H_1}$. In both of these driving
schemes, the state of the system at multiples of the driving period,
$\ket{\psi_n} = \ket{\psi(n T)}$, can be obtained by repeated application of
$U_F$, i.e., $\ket{\psi_n} = U_F^n \ket{\psi_0}$, where $\ket{\psi_0}$ is the
initial state of the system.

We now proceed to discuss how the above driving protocols are modified in the
presence of timing noise. For simplicity, we assume that there is noise only in
the first step of the driving cycle. The generalization of our considerations to
include noise in the second step or to a driving protocol that comprises more
than two steps is straightforward and summarized in the next section. If there
is timing noise in the first step, the Hamiltonian $H_1$ is not applied exactly
for a time $T_1$. Instead, in the $n$-th cycle, it is applied for
$T_1 + \tau_n$, where $\tau_n$ is a random number with zero mean,
$\overline{\tau_n} = 0$ and fluctuations $\overline{\tau_n^2} = \tau^2$.
Strictly speaking, causality requires that $\tau_n \geq -T_1$, but formally we
can relax this constraint, and we simply require that $\tau \ll T_1$. (In
particular, this allows us to take the distribution of the $\tau_n$ to be
Gaussian as in the numerical results presented in the main text.) The precise
form of the probability distribution is not important in the
following. Moreover, we assume that the time shifts in different cycles are
uncorrelated, i.e., $\overline{\tau_n \tau_{n'}} = \tau^2 \delta_{nn'}$.  Under
these conditions, after $n$ driving cycles the state of the system is given by
\begin{equation}
  \ket{\psi_n} = e^{-i T_2 H_2} e^{-i \left( T_1 + \tau_n \right) H_1} \dotsb
  e^{-i T_2 H_2} e^{-i \left( T_1 + \tau_1 \right) H_1} \ket{\psi_0} = U_{F, n}
  \dotsb U_{F, 1} \ket{\psi_0}.
\end{equation}
In the last equality, we introduced the noisy Floquet operator $U_{F, n}$, which
describes the evolution of the system during the $n$-th driving cycle. It can be
written as
\begin{equation}
  \label{eq:UF_binary_noisy}
  U_{F, n} = \mathsf{T} e^{-i \int_{t_{n - 1}}^{t_n} dt \, H(t)} =
  e^{-i T_2 H_2} e^{-i \left( T_1 + \tau_n \right) H_1},
\end{equation}
where $t_n = n T + \sum_{n' = 1}^n \tau_n$ depends on all prior time shifts. The
time-dependent Hamiltonian is now given by
\begin{equation}
  \label{eq:noisy_Hamiltonian}
  H(t) =
  \begin{cases}
    H_1, & t_n \leq t < t_n + T_1 + \tau_n, \\
    H_2, & t_n + T_1 + \tau_n \leq t < t_n + T_1 + \tau_n + T_2.
  \end{cases}
\end{equation}
For the kicking protocol defined by Eq.~\eqref{eq:H_kicked}, the noisy
Hamiltonian assumes the form
\begin{equation}
  \label{eq:H_kicked_noise}
  H(t) = H_1 + \lambda \sum_n \delta(t - t_n) H_2,
\end{equation}
where $t_n$ is defined as above, i.e., the duration between two kicks is
$t_{n + 1} - t_n = T + \tau_n$. The corresponding Floquet operator reads as in
Eq.~\eqref{eq:UF_binary_noisy}, only with $U_2 = e^{-i T_2 H_2}$ replaced by
$e^{-i \lambda H_2}$. In both cases, the evolution of the state
$\ket{\psi_n} = \ket{\psi(t_n)}$ during one driving cycle is given by
\begin{equation}
  \label{eq:psi_evolution}
  \ket{\psi_{n + 1}} = U_{F,n + 1} \ket{\psi_n} = U_2 e^{-i \left( T_1 +
      \tau_{n + 1} \right) H_1} \ket{\psi_n}.
\end{equation}
Evidently, the state of the system at time $t_n$ depends on the particular noise
realization, i.e., on the sequence of all prior time shifts
$\tau_1, \dotsc, \tau_n$. Taking the average over noise realizations, the
expectation value of an observable $O$ can be written as
\begin{equation}
  \label{eq:O}
  \overline{\langle O_n \rangle} = \overline{\braket{\psi_n | O | \psi_n}} = \tr
  \! \left( O \overline{\ket{\psi_n} \bra{\psi_n}} \right) = \tr(O \rho_n),
\end{equation}
where $\rho_n = \overline{\ket{\psi_n} \bra{\psi_n}}$ is the density matrix that
describes the noise-averaged state of the system at time $t_n$. Thus, to
evaluate the expectation values of observables, it is sufficient to track the
evolution of $\rho_n$, and in the following we derive an evolution equation for
this quantity in the weak-noise limit. A key point that facilitates this
derivation is that the noise average can be performed for each driving cycle
individually because the $\tau_1, \dotsc, \tau_n$ are statistically
independent. Thus, rewriting Eq.~\eqref{eq:psi_evolution} for the density
matrix, we obtain
\begin{equation}
  \label{eq:rho_evolution}
  \rho_{n + 1} = \overline{U_{F, n + 1} \rho_n U_{F, n + 1}^{\dagger}}.
\end{equation}
For weak noise in the sense that $\tau \norm{H_1} \ll 1$ where
$\norm{\makebox[1ex]{$\cdot$}}$ is a suitably chosen operator norm, we can
expand the noisy Floquet operator~\eqref{eq:UF_binary_noisy} in the time
shift. Keeping terms up to second order in $\tau_{n + 1}$, we find
\begin{equation}
  \label{eq:UF_expansion}
  U_{F, n + 1} = U_F \left( 1 - i \tau_{n + 1} H_1 - \frac{\tau_{n + 1}^2}{2} H_1^2 \right),
\end{equation}
where $U_F = U_2 U_1$ is the Floquet operator for perfectly periodic
driving. Inserting this form in Eq.~\eqref{eq:rho_evolution}, the average over
noise becomes straightforward, and we obtain (dropping terms of cubic and higher
order in $\tau_{n + 1}$)
\begin{equation}
  \label{eq:discrete_Lindblad_binary}
  \begin{split}
    \rho_{n + 1} & = U_F \overline{\left( 1 - i \tau_{n + 1} H_1 - \frac{\tau_{n
            + 1}^2}{2} H_1^2 \right) \rho_n \left( 1 + i \tau_{n + 1} H_1 -
        \frac{\tau_{n + 1}^2}{2} H_1^2 \right)} U_F^{\dagger} \\ & = U_F
    \overline{\left[ \rho_n - i \tau_{n + 1} \left[ H_1, \rho_n \right] +
        \tau_{n + 1}^2 \left( H_1 \rho_n H_1 - \frac{1}{2} \left\{ H_1^2, \rho_n
          \right\} \right) \right]} U_F^{\dagger} \\ & = U_F \left[ \rho_n - i
      \overline{\tau_{n + 1}} \left[ H_1, \rho_n \right] + \overline{\tau_{n +
          1}^2} \left( H_1 \rho_n H_1 - \frac{1}{2} \left\{ H_1^2, \rho_n
        \right\} \right) \right] U_F^{\dagger} \\ & = U_F \left( \rho_n + \tau^2
      \mathcal{D}[L_1] \rho_n \right) U_F^{\dagger}.
  \end{split}
\end{equation}
This discrete-time evolution equation combines the usual noise-free coherent
stroboscopic Floquet evolution, described by the Floquet operator $U_F$, with
dissipative dynamics as familiar from quantum master equations in Lindblad
form. In particular, in analogy to the usual continuous-time Lindblad equation,
we identify the quantum jump operator $L_1 = H_1$ and the dissipator
\begin{equation}
  \label{eq:dissipator}
  \mathcal{D}[L] \rho = L \rho L - \frac{1}{2} \left\{ L^2, \rho \right\} =
  \frac{1}{2} \left[ \left[ L, \rho \right], L \right].
\end{equation}
In the Floquet-Lindblad equation (FLE)~\eqref{eq:discrete_Lindblad_binary},
these two elements of the evolution --- coherent evolution and dissipation ---
are applied in a staggered fashion, i.e., the map $\rho_n \mapsto \rho_{n + 1}$
is a composition of $\rho \mapsto \rho + \tau^2 \mathcal{D}[L_1] \rho$ and
$\rho \mapsto U_F \rho U_F^{\dagger}$. For comparison, the continuous-time form
of the Lindblad equation reads
\begin{equation}
  \frac{d \rho}{d t} = -i [H, \rho] + \gamma \mathcal{D}[L] \rho.  
\end{equation}
We note that while in the present context of noise in Floquet systems the jump
operator $L_1 = H_1$ is always Hermitian, this is not the case in general. Then,
the dissipator should be modified to
$\mathcal{D}[L] \rho = L \rho L^{\dagger} - \frac{1}{2} \left\{ L^{\dagger} L,
  \rho \right\}$.
In the usual (continuous-time) quantum master equation in Lindblad form, the
dissipator describes the effect of an environment or bath on the system
dynamics, and assumes the above time-local form if (i) the system-bath coupling
is weak and (ii) the bath correlation time is much shorter than the time scales
of the system dynamics (see, e.g.,~\cite{Gardiner2000}). Conditions (i) and (ii)
justify the Born and Markov approximations, respectively, which are made in
derivations of the Lindblad equation. In the derivation of the FLE, the
expansion in the noise strength $\tau$ is analogous to the Born approximation,
and the noise we consider is Markovian (i.e., uncorrelated on the intrinsic time
scale $T$ of the evolution of system) by assumption. Just like the usual master
equation, the FLE can immediately be seen to be trace-preserving and completely
positive.

\subsection{Multi-step piecewise constant driving}
\label{sec:multi-phase-pc}

The above derivation can be generalized straightforwardly to extended driving
protocols, defined in terms of a sequence of Hamiltonians $H_i$ with
$i = 1, 2, \dotsc, M$ which are applied for times $T_i$ so that the duration of
a full driving cycle is $T = \sum_{i = 1}^M T_i$. Assuming --- as in the main
text --- that there is timing noise in each step of the driving cycle, the FLE
that generalizes Eq.~\eqref{eq:discrete_Lindblad_binary} takes the form
\begin{equation}  
  \label{eq:discrete_Lindblad}
  \rho_{n + 1} = U_F \left( \rho_n + \tau^2 \sum_{i = 1}^M \mathcal{D}[L_i]
    \rho_n \right) U_F^{\dagger},
\end{equation}
where $U_F = U_M \dotsb U_1$ with $U_i = e^{-i T_i H_i}$. Here we take the
shifts $\tau_{ni}$ in different steps and driving cycles to be uncorrelated and
identically distributed,
$\overline{\tau_{ni} \tau_{n' i'}} = \tau^2 \delta_{nn'} \delta_{ii'}$. The jump
operators are given by
\begin{equation}
  \label{eq:L}
  L_i = U_F^{\dagger} U_M \dotsb U_i H_i U_{i - 1} \dotsb U_1 = U_1^{\dagger}
  \dotsb U_{i - 1}^{\dagger} H_i U_{i - 1} \dotsb U_1.
\end{equation}
As above, the jump operators are Hermitian, $L_i^{\dagger} = L_i$.

\subsection{Reduction to classical master equation}
\label{sec:reduct-class-master}

In many cases of practical interest, the discrete Lindblad equation for the
density matrix can be reduced to a classical master equation for the diagonal
elements of the density matrix written in the basis of Floquet eigenstates
$\ket{\alpha}$. The latter are the right eigenvectors of the Floquet operator,
$U_F \ket{\alpha} = e^{-i T \epsilon_{\alpha}} \ket{\alpha}$, and
$\epsilon_{\alpha}$ is the respective quasienergy. In this basis, the density
matrix can be written as
\begin{equation}  
  \rho = \sum_{\alpha, \beta} \rho^{\alpha \beta} \ket{\alpha}
  \bra{\beta}. 
\end{equation}
The condition for obtaining a classical master equation is that upon repeated
application of $U_F$ in Eq.~\eqref{eq:discrete_Lindblad}, the off-diagonal
elements of the density matrix dephase more rapidly than they are repopulated
from the diagonal. Then, for the diagonal elements
$\rho^{\alpha} = \rho^{\alpha \alpha}$ we obtain the discrete master equation
\begin{equation}
  \label{eq:master_equation}
  \rho^{\alpha}_{n + 1} =  \sum_{\beta} W_{\beta \to \alpha} \rho^{\beta}_n,
\end{equation}
where the transition probabilities are given by
\begin{equation}
  \label{eq:transition_probabilities}
  W_{\beta \to \alpha} = \delta_{\alpha \beta} + \tau^2 \sum_i \left( \abs{\braket{\alpha | L_i |
        \beta}}^2 - \braket{\alpha | L_i^2 | \alpha} \delta_{\alpha \beta}
  \right).
\end{equation}
The transition probabilities are symmetric,
$W_{\alpha \to \beta} = W_{\beta \to \alpha}$, and conservation of probability
--- in other words, conservation of the trace of the density matrix --- results
in $\sum_{\beta} W_{\beta \to \alpha} = 1$. Thus, the probability to remain in
state $\alpha$ can be written as $W_{\alpha \to \alpha} = 1 - \sum_{\beta \neq
  \alpha} W_{\beta \to \alpha}$, and Eq.~\eqref{eq:master_equation} can be
recast as
\begin{equation}  
  \rho^{\alpha}_{n + 1} = \rho^{\alpha}_n + \sum_{\beta \neq \alpha} W_{\beta
    \to \alpha} \left( \rho^{\beta}_n - \rho^{\alpha}_n \right),
\end{equation}
which is the form quoted in the main text.

\section{Decay of an End-State in a Noisy Floquet Topological Chain}

Here, we apply the formalism derived above to derive the time evolution of an
imperfectly driven ladder as introduced in the main text. We show that timing
noise leads to diffusion exactly on resonance and to exponential decay away from
resonance. The microscopic parameter entering the discussion is the hopping
parameter $J = 2\pi/T + \delta J$, with $\delta J$ being a measure for the
bulk's bandwidth. We assume that $\delta J$ is small, enabling us to make
analytical progress. In the following, we will distinguish the point of resonant
driving $\delta J=0$ from the limit $\delta J \to 0$ and show how they capture
the diffusive and exponential decay, respectively. Working in the eigenbasis of
the Floquet operator, we find that transitions between Floquet states are
induced by the noise at order $J^2\tau^2$. For small deviations from resonance,
in the calculation of the transition probabilities we can set $\delta J = 0$
(thus, ignoring terms $O(J^2\tau^2 \delta J^2)$). The crucial point is that we
have to be careful in taking limit $\delta J \to 0$: starting from finite
$\delta J$ and sending it to zero, the bulk states remain delocalized with an
arbitrarily small but finite quasienergy and bandwidth; on the other hand,
working \emph{exactly} on resonance, we should take the bulk states to be
localized with an exact degeneracy of all states at quasienergy $0$.

Employing the formalism of the Floquet-Lindblad equation derived above, we note
that the periodic drive considered in the main text comprises $M = 4$ individual
steps. Thus, the Floquet operator is $U_F = U_4 U_3 U_2 U_1$, with the
individual steps $U_i = e^{-iT H_i/4}$ where
$H_i = -\sum_{\mu \nu}J_{\mu \nu}^i \left( c^\dagger_{\mu} c_{\nu} + \Hc
\right)$,
as introduced in the main text. The time evolution during one step can thus be
evaluated assuming single particle states and using the fact that the sum over
bonds in $H_i$ is over mutually disconnected pairs of neighboring lattice
sites. Using for $n \geq 1$ the identities
\begin{align}
  H_i^{2n} = J^{2n} \sum_{\braket{\mu \nu}_i} \left( n_\mu + n_\nu \right)
  \qquad \text{and} \qquad
  H_i^{2 n - 1} = J^{2 \left( n - 1 \right)} H_i,
\end{align}
with the sum running over all bonds affected by the hopping in step $i$, we find
\begin{align}
	U_i = 1 + \left( \cos(\phi) - 1 \right) \sum_{\braket{\mu \nu}_i} (n_\mu
  + n_\nu) - i \sin(\phi) \, H_i/J,
\end{align}
with $\phi = JT/4$. Equivalently, we can represent the $U_i$ in the single
particle basis with the projector on all involved lattice sites
$P_i = \sum_{\braket{\mu \nu}_i} \left(\ket{\mu}\bra{\mu} +
  \ket{\nu}\bra{\nu}\right)$ and its complement $Q_i = 1 - P_i$:
\begin{align}
	U_i = P_i \left(\cos(\phi) - i \sin(\phi) \, H_i /J\right) P_i + Q_i ,
\end{align}
as used in the main text. In steps $1$ and $3$ all lattice sites are involved
and $Q_1=Q_3=0$, while in steps $2$ and $4$ one site at each end drops out of
the respective Hamiltonian and thus
$Q_2 = \ket{1, + }\bra{1, + }+\ket{L-1,- }\bra{L-1,-}$ and
$Q_4 = \ket{0,- }\bra{0,- }+\ket{L,+ }\bra{L,+}$.  The jump operators $L_i$
introduced above are $L_1 =H_1$, $L_2 = U_1^{\dagger}H_2 U_1$,
$L_3 =U_1^{\dagger} U_2^{\dagger}H_3 U_2 U_1$, and
$L_4 =U_1^{\dagger} U_2^{\dagger} U_3^{\dagger}H_4 U_3 U_2 U_1$.

The form of the jump operators simplifies further exactly on resonance, when
$\phi = \pi/2$ and during each step of the driving protocol a particle is fully
transferred from one lattice site to another. Then, we find $L_1 =H_1$,
$L_2=H_1H_2H_1$,
$L_3=H_1 \left( H_2-iQ_2 \right) H_3 \left( H_2+iQ_2 \right)H_1$, and
$L_4=H_1 \left( H_2-iQ_2 \right) H_3H_4H_3 \left( H_2+iQ_2 \right)H_1$.
Moreover, on resonance the Floquet eigenstates take a particularly simple form:
The end states are localized to the outermost lattice sites
$\ket{e_l} = \ket{0,-}$ and $\ket{e_r}=\ket{L,+}$, while for the bulk states we
can choose the lattice site basis $ \ket{b} = \ket{j,s}$ with $j=1,\dots,L-1$
and $s=\pm$. In the lattice basis, the jump operators are represented as
$L_1 = J \sum_{j} \left( \ket{j,+}\bra{j-1,-} + \Hc \right)$,
$L_2 = J \sum_{j} \left( \ket{j,+}\bra{j,-} + \Hc \right)$,
$L_3 = J \sum_{j} \left( \ket{j,+}\bra{j+1,-} + \Hc \right)$, and
$L_4 = J \sum_{j} \left( \ket{j,+}\bra{j,-} + \Hc \right)$. Focusing on the left
end state $\ket{e} \equiv \ket{e_l}$, we immediately see that the diagonal
matrix elements of the jump operators vanish, $\braket{e|L_i|e} = 0$, and that
the left end state is an eigenstate of the square of the jump operators:
\begin{equation}  
  \label{eq:left_edge_L2}
  L_i^2 \ket{e} =
  \begin{cases}
    J^2 \ket{e}, & i=1,3, \\
    0, & i = 2,4.
  \end{cases}
\end{equation}
This, in turn, implies that $\braket{e|L_i^2|e}=J^2$ for $i=1,3$ and
$\braket{e|L_i^2|e}=0$ otherwise, while $\braket{b | L_i^2 | e} = 0$. Below, we
investigate the stability of the chain's end state very close to as well as
exactly on resonance. In both cases, we can use the above expressions for matrix
elements of the jump operators which we derived for resonant driving. The
crucial difference comes from the matrix elements $\braket{b | L_i | e}$, which
we discuss in detail below.

In the following, we restrict our attention to the left end of the chain which
we denote by $\ket{e}$, and we disregard the other one, assuming a semi-infinite
system. Then, the system's density matrix can be represented in the Floquet
eigenbasis as
\begin{align}
  \rho_n = \rho_n^e \ket{e}\bra{e} 
  + \sum_b \left(\rho_n^{eb} \ket{e}\bra{b} + \rho_n^{be} \ket{b}\bra{e}  \right) 
  +\sum_{bb'}\rho_n^{bb'}\ket{b}\bra{b'} .
\end{align}
Evolving the density matrix by one driving period with the
FLE~\eqref{eq:discrete_Lindblad}, we find for the occupation of the chain's left
end state:
\begin{multline}
  \rho^e_{n+1} = \left(1 + \tau^2 \sum_i \left( \left| \braket{e|L_i|e}
      \right|^2 - \braket{e|L_i^2|e}\right) \right) \rho^{e}_n \\ + \tau^2
  \sum_i \sum_b \left[ \left( \braket{e | L_i | e} \braket{b | L_i | e} -
      \frac{1}{2} \braket{b | L_i^2 | e} \right) \rho_n^{eb} + \left( \braket{e
        | L_i | b} \braket{e | L_i | e} - \frac{1}{2} \braket{e | L_i^2 | b}
    \right) \rho_n^{be} \right] \\ + \tau^2 \sum_i \sum_{bb'} \braket{e|L_i|b}
  \braket{b'|L_i|e} \rho^{bb'}_n.
\end{multline}
This can be simplified using the matrix elements evaluated above, and we obtain
(cf.\ Eq.~(5) of the main text)
\begin{align}
  \label{eq:evo_rho_exp}
  \rho^{e}_{n+1} 
  = \left(1- 2 \tau^2 J^2 \right) \rho^{e}_n
  + \tau^2 \sum_i \sum_{bb'} \braket{e|L_i|b} \braket{b'|L_i|e} \rho^{bb'}_n,
\end{align}
where, in particular, coherences between the end state and bulk states drop out.
    
\subsection{Exponential Decay for a Dispersive Bulk} 
\label{sub:exponential_decay}

We first analyze the noisy stroboscopic time evolution of the end states in
Eq.~\eqref{eq:evo_rho_exp} for a dispersive bulk, i.e., $\delta \phi \neq 0$,
for which the bulk states acquire a finite band width $\propto \delta \phi$ and
are delocalized over the full length of the system. An analytical treatment is
possible only close to resonant driving, and hence we focus on the limit
$\delta \phi \to 0$. This allows us to approximate the end state $\ket{e}$ and
the jump operators with their forms at resonant driving. Importantly, the bulk
states have finite albeit arbitrarily small quasienergies.

For the matrix elements connecting the end to the bulk states, $\braket{b|L_i|e}$,
we need to keep in mind that the jump operators transfer electrons only between
lattice sites within a short range of each other. Assuming that the bulk states
are evenly spread out over the entire chain, they carry a normalization factor
of $\sim 1/\sqrt{L}$, leading to the following scaling of the matrix elements
with system size:
\begin{align}
  \braket{b|L_i|e}\sim {1\over \sqrt{L}} .
\end{align}

Moreover, it is important to note that in the delocalized basis the off-diagonal
part of the bulk density matrix $\rho_n^{bb'}$ for $b\neq b'$ is non-zero. These
matrix elements are getting populated by excitations from the end state at a
rate $\tau^2 \braket{b|L_i|e}\braket{e|L_i|b'}\sim \tau^2/L$ and, to leading
order, evolve coherently with $U_F$ per period. The off-diagonal elements
therefore pick up phase factors of $e^{i T \left( \varepsilon_b-\varepsilon_b' \right)}$
for each full period. No matter how small the quasienergies, these phases
accumulate and average to zero for long enough times. Only the diagonal elements
are static and contribute to the sum over bulk states in
Eq.~\eqref{eq:evo_rho_exp}, which then decays with the system size as can be
seen from a simple dimensional analysis
\begin{align}
	\sum_{b} \rho^{bb}_n \left|\braket{e|L_i|b}\right|^2 \propto {1\over L} .
\end{align}
For large enough systems and long enough times, the contribution from this sum
can be neglected and leaves us with an exponential decay of the end state (cf.\
Eq.~(6) of the main text)
\begin{align}
  \rho^{e}_{n+1} 
  = \left(1- \tau^2 J^2 \right) \rho^{e}_n + O(1/L),
\end{align}
confirming the intuitive picture that a dispersive bulk would carry away any
excitation out of the edge. This contrasts the case of the localized bulk at
resonant driving in which excitations get stuck close to the end state and have
a finite return probability, as we discuss in the next section.

\subsection{Diffusive Decay at Resonant Driving} 
\label{sub:diffusive_decay_at_resonant_driving}

We now focus on the dynamical evolution of the density matrix for a flat bulk
band at resonant driving. As explained above, in this case we can work in the
basis of lattice sites, in which the end states are
$\ket{e_l} = \ket{0,-} \equiv \ket{e}$ and $\ket{e_r}=\ket{L,+}$, while the bulk
states are give by $ \ket{b} = \ket{j,s}$ with $j=1,\dots,L-1$ and $s=\pm$. Note
that the lattice site basis for the bulk is an arbitrary choice since the bulk
band is fully degenerate.

In the equation for the end state population~\eqref{eq:evo_rho_exp}, only $L_1$
and $L_3$ connect the end state to the bulk states, such that
$\braket{e|L_i|b}=J$ for $i=1,3$ and $\ket{b_1}=\ket{1,+}$ and
$\ket{b_3}=\ket{1,-}$. All other elements $\braket{e|L_i|b}$ are zero. Then, the
dynamical equation reduces to
\begin{align}
  \label{eq:edge_diffusive}
	\rho^{0,-}_{n+1} =
	\left(1 - 2 \tau^2 J^2\right) \rho^{0,-}_{n} 
	+ \tau^2 J^2 \left( \rho^{1,+}_{n} + \rho^{0,-}_{n}\right) ,
\end{align}
where we introduced a notation for the density matrix diagonal labeled by
lattice sites $\rho_n^{bb}=\rho_n^{j,s}$. We note that all off-diagonal elements
of the density matrix drop out of Eq.~\eqref{eq:edge_diffusive} --- a property
that is also true for the evolution equation of the population of a generic bulk
state. Thus, starting from an initially diagonal density matrix,
$\rho_n = \ket{e} \bra{e}$, the density matrix remains diagonal at all
times. Importantly, the time evolution for the end state is strongly coupled to
its neighboring bulk states. To understand its behavior in the long-time limit
we can alternatively look at the dynamical equation for a generic bulk state,
\begin{equation}
  \begin{split}
  \rho^{j,s}_{n+1} &= \left(1 - \tau^2 \sum_i \sum_{l,s'}
                     \braket{l,s'|L_i|j,s}^2 \right) \rho^{j,s}_n + \tau^2 \sum_i \sum_{l,s'}
                     \braket{l,s'|L_i|j,s}^2 \rho^{l,s'}_n
                      \\
                   &=  \left(1 - 4\tau^2 J^3\right) \rho^{j,s}_n
                     + \tau^2 J^2 \left( 2 \rho_n^{j,-s} + \rho_n^{j+s,-s} +
                       \rho_n^{j-s,-s} \right).
\end{split}
\end{equation}
The last equation is Eq.~(7), and can straightforwardly be reduced to the
diffusion equation~(8) for the total occupation of doublets
$\rho^j_n = \rho^{j, +}_n + \rho^{j,-}_n$. As bulk and end state can be treated
in the same way at resonant driving, being localized to one lattice site each,
the behavior of the end state for long times can be inferred immediately. It
obeys a diffusion equation against a hard wall boundary for a particle
initialized right next to this boundary.

Note that also in the resonant driving case we can choose to describe the bulk
in a delocalized basis, rendering the off-diagonal elements $\rho_n^{bb'}$
non-zero. However, in contrast to the dispersive case, these elements are static
in the stroboscopic time-evolution due to the flat bulk band and can thus not be
neglected.

\section{Numerical simulations and disorder effects}

Adding a chemical potential disorder to the Floquet system breaks particle-hole symmetry, such that the end modes are no longer topologically protected. However, since in the clean case the end modes are well localized at the ladder boundaries and separated in quasienergy from bulk states, one can expect these features to persist for small enough disorder strength. We show here that this is indeed the case, by determining the average localization lengths and positions of Floquet eigenstates in the presence of on-site disorder. On a disordered ladder consisting of $L=200$ rungs, we compute the real space probability distribution of each Floquet eigenstate, $|\psi(x)|^2$ with $x=1,\ldots, L$ indexing the rungs of the ladder, and find the associated localization length by fitting with an exponential decay of the form
\begin{equation}\label{eq:loclen}
 f(x) = A \exp \left( -\frac{|x-x_0|}{\xi} \right).
\end{equation}
The localization length $\xi$ is extracted from the fit, while the amplitude $A$ and the wavefunction center $x_0$ are taken to be the value and position of the maximum of each $|\psi(x)|^2$. From the values of $x_0$ we also determine the average displacement of states from the center of the chain, $d=|x_0 - L/2|$. Since $x_0 \in [1,L]$, the displacement takes values $d\in [0, L/2]$. States located on the boundaries of the system are therefore expected to show a large displacement $d\to L/2$, while states that are uniformly distributed throughout the bulk should have an average displacement $d\to L/4$.

\begin{figure}[tb]
 \includegraphics[width=0.9\textwidth]{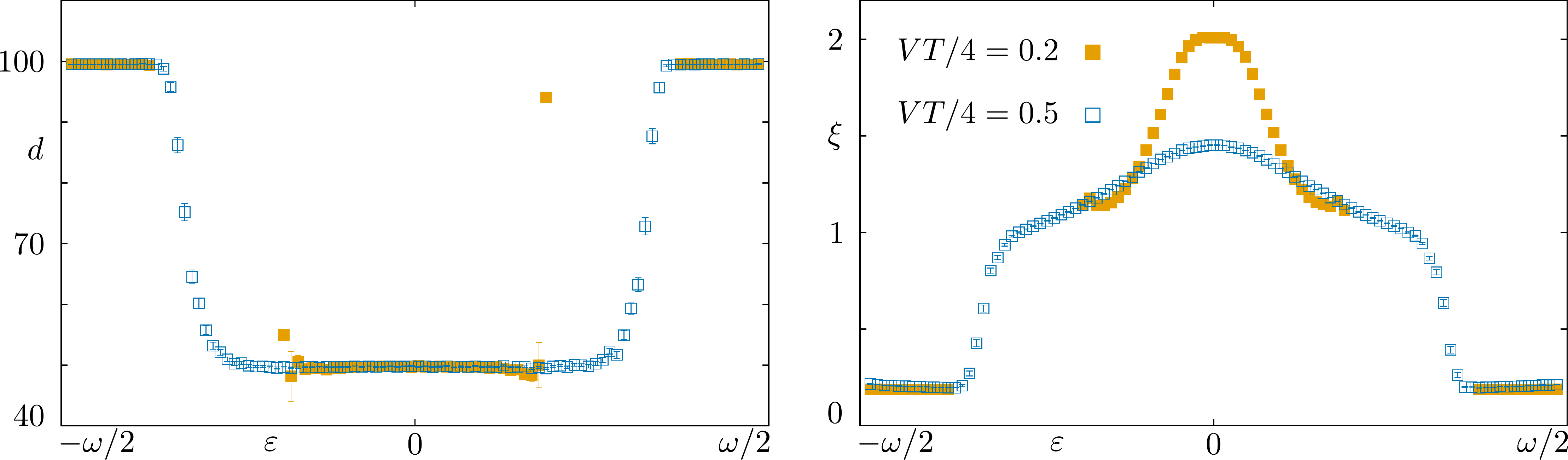}
 \caption{Average displacement $d$ (left) and localization length $\xi$ (right) of Floquet eigenstates on ladder of $L=200$ rungs, using $\phi=1.45$ as in the main text Fig. 2. The on-site disorder strength is set to $VT/4=0.2$ (solid orange squares) and $VT/4=0.5$ (empty blue squares). Each point represents an average over states in a range of quasienergies centered at $\varepsilon$ and having a width of $\omega/101$. A total of $10^4$ independent disorder realizations were used for each disorder strength. As disorder strength is increased, bulk states around $\varepsilon=0$ spread in quasienergy and become more localized, but remain uniformly distributed throughout the system, showing an average displacement $d\simeq L/4=50$. Edge modes close to $\varepsilon=\pm\omega/2$ show a much smaller localization length as compared to bulk states, and remain localized at the ends of the system, with an average displacement $d\simeq L/2=100$. \label{fig:loclen}}
\end{figure}

The results shown in Fig.~\ref{fig:loclen} confirm our expectations. States with quasi energies close to $\varepsilon=0$ are uniformly distributed throughout the chain. They become more localized and spread out in quasienergy with increasing disorder strength.
However, states close to $\varepsilon=\pm\omega/2$ are localized exclusively on the chain boundaries and have a much smaller localization length as compared to bulk states. As such, even with a small on-site disorder strength, edge modes remain well localized on the system boundaries and separated in quasi-energy from the bulk states.

\begin{figure}[tb]
 \includegraphics[width=0.4\textwidth]{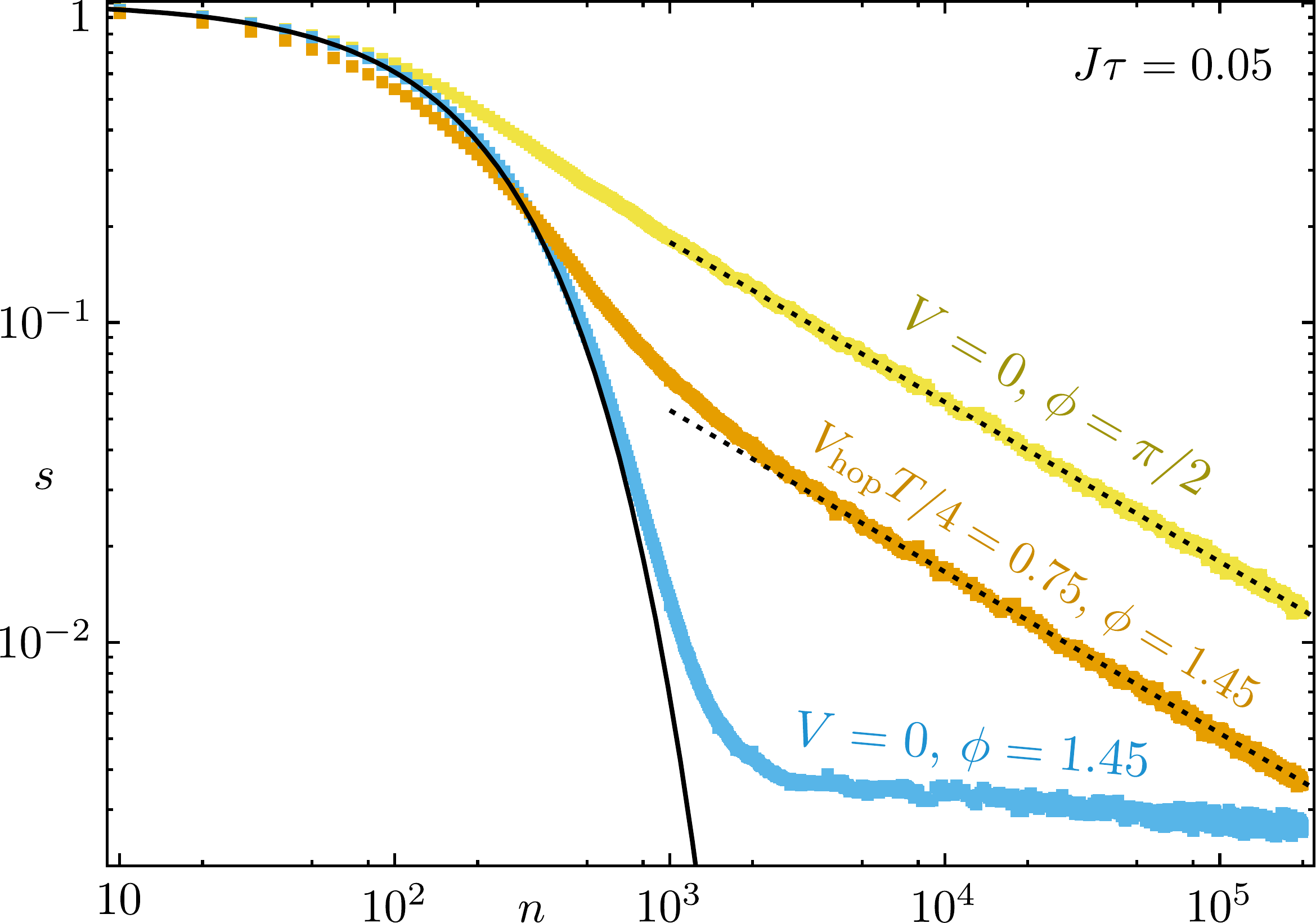}
 \caption{As in the main text, we plot the survival probability of an initial
   end mode, $s=\overline{\braket{e | \psi_n}^2}$ as a function of the
   number of noisy driving cycles, $n$. All system parameters, as well as the
   blue and yellow curves are the same as in the main text Fig.~2. The orange
   curve however is obtained by introducing disorder in the hopping amplitudes,
   using $V_{\rm hop}T/4=0.75$, and averaging over 4000 independent realizations
   of disorder and noise. Since hopping disorder also localizes all bulk states,
   the decay is again diffusive, as indicated by the dashed
   line.\label{fig:1dhopping}}
\end{figure}

As a final point, we show numerically that hopping
disorder also slows the end mode decay down to a diffusive process. This is
indicated in Fig.~\ref{fig:1dhopping}, and is a consequence of the fact that
hopping disorder in this model leads to a localization of all bulk states. While
localization is expected for on-site disorder, which breaks the particle-hole
symmetry, this is not immediately obvious in the case of random hoppings, since
the particle-hole symmetry of the Floquet operator, Eq.~\eqref{eq:FPHS}, is
preserved for every disorder realization. One-dimensional particle-hole
symmetric chains may enter a so-called critical phase, characterized by the
presence of delocalized states \cite{Brouwer2000, Motrunich2001, Brouwer2003,
  Gruzberg2005}. In the 1D ladder model, however, Floquet bulk bands form a
parabolic (as opposed to linear) band touching point at the particle-hole
symmetric quasienergy $\varepsilon=0$, meaning that the bulk state velocity
vanishes already in the clean limit. In this way, the 1D critical phase is
avoided, all bulk states become localized, and the decay is diffusive as in the
case of on-site disorder.

\bibliography{bibliography}